# The OSU Scheme for Congestion Avoidance in ATM Networks: Lessons Learnt and Extensions[1]

Raj Jain, Shiv Kalyanaraman and Ram Viswanathan[2]
Computer and Information Sciences, The Ohio State University
395 Dreese, 2015 Neil Ave., Columbus, OH 43210-1277
E-mail: {*jain, shivkuma*}@cis.ohio-state.edu, ramv@microsoft.com

## Abstract

The OSU scheme is a rate-based congestion avoidance scheme for ATM networks using explicit rate indication. This work was one of the first attempts to define explicit rate switch mechanisms and the Resource Management (RM) cell format in Asynchronous Transfer Mode (ATM) networks. The key features of the scheme include explicit rate feedback, congestion avoidance, fair operation while maintaining high utilization, use of input rate as a congestion metric, O(1) complexity. This paper presents an overview of the scheme, presents those features of the scheme that have now become common features of other switch algorithms and discusses three extensions of the scheme.

## 1 Introduction

The amount of data that can be lost due to congestion on a link depends upon its delay-bandwidth product. On high-speed networks, this amount can be large and so it is particularly important to have proper congestion control in such networks [4, 12, 15]. This is why traffic management is such a hot topic in ATM Forum, International Telecommunications Union (ITU), and Internet Engineering Task Force (IETF), where high-speed networking standards of tomorrow are being designed.

Until about 1994, most of the congestion control work was based on window flow controls (e.g., in TCP/IP). Feedback was either implicit (e.g., via timeouts in TCP/IP [5]) or explicit but binary (e.g., in DECbit [8] or its derivatives). In fact, until July 1994, even ATM networking standards used an explicit binary feedback method called "Explicit Forward Congestion Indication (EFCI) [6]. The ATM Forum then decided to include an "explicit rate" approach in which the switches tell the sources the exact rate that they can use [1].

In the basic approach decided by the Forum, the sources periodically send resource management (RM) cells that contain, among other things, indication of a source's current load. The RM cells travel through the network to the destination, which then returns them to the source. The switches provide the feedback to the sources by overwriting appropriate fields of these RM cells.

This paper presents results of our first attempt to design an explicit rate scheme for ATM networks. This work was done between July and October 1994. The scheme is a follow on to the MIT scheme [1], and hence is named the Ohio State University (OSU) scheme. This research has helped us and the Forum understand many issues that were not so well understood before and formulate approaches for tackling these issues. The purpose of this paper is to document the issues, present our approaches to resolve these issues, and our results using a few schemes based on these approaches. The OSU scheme was the first of these schemes.

An overview of the OSU scheme was published recently in a workshop [7]. We first present a summary of the scheme in section 2. In section 3, we present the lessons learnt from this work and the key contributions of this research that either have become commonly accepted parts of switch schemes or have been adopted by the standard. We then present three extensions of the basic OSU scheme in section 4, and sample simulation results in section 5. The direction of ATM forum has changed considerably as the scheme was in development. Therefore, some the features of this scheme have to be changed to be compatible with current ATM Forum standards. These changes and their impact are described in section 6. We then discuss the limitations of the scheme and summarize the paper in section 7. Appendix A gives a proof for the fairness algorithm and appendix B gives the detailed pseudocode for all the algorithms.

---

# 2 OSU Scheme: An Overview

## 2.1 The Source Behavior

The OSU scheme requires sources to send RM cells **periodically** at intervals of $T$ microseconds. The RM cells contain several fields. Three of these are: transmission cell rate (TCR), the average offered cell rate (OCR), and a load adjustment factor (LAF). The TCR is the inverse of the minimum inter-cell transmission time and indicates instantaneous peak load put by the source. For bursty sources, TCR is not a good indication of overall load. Therefore, the average load over $T$ interval is indicated in the OCR field of the RM cell. Normally the instantaneous peak rate (TCR) is more than the average rate (OCR). However, when TCR has just been reduced, the OCR may have a value between the old TCR and current TCR. Hence, we set:

$$\text{TCR in RM Cell} \leftarrow max\{TCR, OCR\}$$

The LAF is the feedback from the network. It indicates the factor by which the source should increase (or decrease) its load. At the source, the LAF is initialized to zero. Switches on the path can only increase the factor. This ensures that successive switches only reduce the rate allowed to the source. Thus, the source receives the rate allowed by the bottleneck along the path.

The source modifies its TCR using the LAF and TCR in the RM cell as follows:

$$\text{New TCR} \leftarrow \frac{\text{TCR in Cell}}{\text{LAF in Cell}}$$

$$\text{IF (LAF} \geq 1 \text{ and New TCR} < \text{TCR)} \; THEN \; \text{TCR} = \text{New TCR}$$

$$\text{ELSE IF (LAF} < 1 \text{ and New TCR} > \text{TCR)} \; \text{TCR} = \text{New TCR}$$

The last two conditions ensure that the source does not inadvertently increase (or decrease) its rates when the network is asking it to decrease (or increase). When LAF $\geq$ 1, the network is asking the source to decrease its TCR. If New TCR is less than the current TCR, the source reduces its TCR to New TCR. No adjustments are required otherwise. The other case (LAF $<$ 1) is similar.

## 2.2 The Switch Behavior

In the OSU scheme, switches compute the feedback when an RM cell is seen in the forward direction. The switch uses *the OCR* (rather than the TCR) in the RM cell for its computation. The OCRs are additive i.e., the sum of OCRs equals the total input load (assuming that the sources are bottlenecked at this switch). Note that the sum of the TCRs may be greater than the total input load. The use of OCRs instead of TCRs in the switch computation allow switches to correctly allocate the rates to the sources.

The OSU scheme is a *congestion avoidance scheme*, that is, it keeps the network at high throughput and low delay in the steady state [8]. For rate-based schemes, the system will be in this region when the sum of the rates of all sources is less than 100 % in the steady state. We use a target utilization ($U$) variable which is set to a fraction (close to, but less than 100 %) of the available capacity. This allows the scheme to achieve high utilization and low queues in steady state. Note that target utilization is set to a value less than 100%. A lower target utilization reduces utilization in the steady state, but reserves more capacity to drain out queues built up due to transient overloads, and vice versa.

The switch measures its current load level, $z$, as the ratio of its "input rate" to its "target output rate". The input rate is measured by counting the number of cells *received* by the switch during a fixed averaging interval.

$$\text{Target Output Cell Rate} = \frac{\text{Target Utilization (U)} \times \text{Link bandwidth in Mbps}}{\text{Cell size in bits}}$$

$$z = \frac{\text{Number of cells received during the averaging interval}}{\text{Target Output Cell Rate} \times \text{Averaging Interval}}$$

The current load level $z$ is used to detect congestion at the switch and determine an overload or underload condition.

### 2.2.1 Achieving Efficiency

To achieve efficiency, the switch replaces the load adjustment factor (LAF) in each RM cell by the maximum of the the current load level $z$ and the LAF value already in the cell.

$$\text{LAF} \leftarrow \max(\text{LAF in the cell}, z) \qquad (1)$$

The idea behind this step is that, if all sources divide their rates by this factor in the current cycle (round trip), the bottleneck link (the link with the maximum utilization) will reach a load level of 1 in the next cycle. This statement is true if all the round trip times are equal and the sources get feedback at the same time (synchronous operation). Otherwise, the bottleneck moves towards a load level of 1 in every cycle, given that sources can use their allocations to send data.

### 2.2.2 Achieving Fairness

Observe that, though the bottleneck reaches a load level of 1, the allocation of the available bandwidth among contending sources may not be fair. This is because, for $z = 1$, the switch does not ask sources to change their rates, even if the distribution of rates is unfair.

Our first goal is to achieve efficient operation. Once the network is operating close to the target utilization ($z = 1$), we take steps to achieve fairness.

For fairness, the network manager declares a target utilization band (*TUB*), say, 90±9% or 81% to 99%. When the link utilization is in the TUB, the link is said to be operating efficiently. The TUB is henceforth expressed in the U(1±$\Delta$) format, where $U$ is the target utilization and $\Delta$ is the half-width of the TUB. For example, 90±9% is expressed as $90(1 \pm 0.1)$%.

We first define a FairShare variable as:

$$\text{FairShare} = \frac{\text{Target Output Cell Rate}}{\text{Number of Active Sources}}$$

A source is said to be active if any cells from the source are seen at the switch during the current averaging interval. To achieve fairness, we need to treat the underloading and overloading sources differently. Underloading sources are those that are using bandwidth less than the FairShare and overloading sources are those that are using more than the FairShare.

If the current load level is $z$, the underloading sources are treated as if the load level is $z/(1 + \Delta)$ and the overloading sources are treated as if the load level is $z/(1 - \Delta)$.

IF (OCR in cell < FairShare) LAF in cell $\leftarrow$ Max(LAF in cell, $\frac{z}{(1+\Delta)}$)
 ELSE LAF in cell $\leftarrow$ Max(LAF in cell, $\frac{z}{(1-\Delta)}$)

As proven in Appendix A, this algorithm guarantees that the system converges towards fair operation. Also, once the bottleneck is inside the TUB, the network remains in the TUB unless the number of sources bottlenecked at this switch changes or their load pattern changes. In other words, TUB is a "closed" operating region. These statements are true for any value of $\Delta$ less than 0.5.

If $\Delta$ is small, as is usually the case, division by $1 + \Delta$ is approximately equivalent to a multiplication by $1 - \Delta$ and vice versa. Note that, a narrow TUB slows down the convergence to fairness (since the formula depends on $\Delta$) but has smaller oscillations in steady state. A wide TUB results in faster progress towards fairness, but has more oscillations in steady state. The size of the TUB is required to be small as indicated in appendix A.

We note that all the switch steps are O(1) w.r.t. the number of VCs. This is an improvement over the M.I.T. scheme, which has a computational complexity of O(n). The detailed pseudo code of the OSU scheme (called the basic fairness option) is given in appendix B.

# 3 Key Contributions of The OSU Scheme Research

The OSU scheme was presented to the ATM Forum traffic management working group in its September and October 1994 meetings. It highlighted several new ideas that have now become common features of most such schemes developed since then.

## 3.1 Use Input Rate Rather Than Queue Length As Congestion Indicator

Most congestion control schemes for packet networks in the past were window based. Most of these schemes use queue length as the indicator of congestion. Whenever the queue length (or its average) is more than a threshold, the link is considered congested. This is how initial rate-based scheme proposals were also being designed. We argued that the queue length is not a good indicator of load when the control is rate-based. With rate-based control, the input rate is a better indicator of congestion. If the input rate is lower than available capacity, the link is not congested even if the queue lengths are high because the queue will be decreasing. Similarly, if the input rate is higher than the available capacity, the system should start taking the steps to reduce congestion since the queue length will be increasing.

Monitoring input rates not only gives a good indication of load level, it also gives a precise indication of overload or underload. For example, if the input rate to a queue is 20 cells per second when the queue server can handle only 10 cells per second, we know that the queue overload factor is 2 and that the input rate should be decreased by a factor of 2. No such determination can be made based on instantaneous queue length. The input rate can hence be used as a metric to compute the new rate allocations. The use of input rates as a metric avoids the use of unnecessary parameters.

Many switch algorithms today use input rate as the congestion indicator.

## 3.2 Use Target Utilization for Congestion Avoidance

Congestion avoidance is distinguished from congestion control by the fact that it allows the network to operate at high throughput and low delay. DECbit and many of its derivative schemes achieve congestion avoidance by trying to keep average queue length at one [8]. With rate-based control, the network will not be overloaded as long as the link utilization is below 100%. Thus, congestion avoidance can be achieved simply by trying to keep the link utilization at a value close to, but below 100%. This is what we call "Target Utilization." This term has now become standard and is used in many other switch algorithms to achieve congestion avoidance.

## 3.3 Use Measured Rather Than Declared Loads

Many schemes prior to OSU scheme, including the MIT scheme, used source declared rates for computing their allocation without taking into account the actual load at the switch. In the OSU scheme, we measure the current total load. All unused capacity is allocated to contending sources. We use the source's declared rate to compute a particular sources' allocation but use the switch's measured load to decide whether to increase or decrease. Thus, if the sources lie or if the source's information is out-of-date, our approach may not achieve fairness but it still achieves efficiency. Again, measuring the total load has become minimum required part of most switch algorithms. Of course, some switches may measure each individual source's cell rate rather than relying on the information in the RM cell.

## 3.4 Count the Number of Active Sources

The OSU scheme introduced the concept of averaging interval and active sources. Most of the virtual circuits (VCs) in an ATM network are generally idle. Its the number of active VCs rather than the total number of VCs that is meaningful. We compute use the number of active VCs to compute fairshare. As discussed in section 7, if the measured value is wrong (which is possible if the averaging interval is short), fairness may not be achieved.

## 3.5 Use Bipolar Feedback

A network can provide two kinds of feedback to the sources. Positive feedback tells the sources to increase their load. Negative feedback tells the sources to decrease their load. These are called two polarities of the feedback.

Some schemes use only one polarity of feedback, say positive. Whenever, the sources receive the feedback, they increase the rate and when they don't receive any feedback, the network is assumed to be overloaded and the sources automatically decrease the rate without any explicit instruction from the network. Such schemes send feedback only when the network is underloaded and avoid sending feedback during overload. The PRCA scheme [13] is an example of a unipolar scheme with positive polarity only.

Unipolar schemes with negative polarity are similarly possible. Early versions of PRCA used negative polarity in the sense that the sources increased the rate continuously unless instructed by the network to decrease. The slow start scheme used in TCP/IP is also an example of unipolar scheme with negative polarity although in this case the feedback (packet loss) is an implicit feedback (no bits or control packets are sent to the source).

The OSU scheme uses both polarities. The DECbit scheme [8] is another example of a bipolar scheme. Since current ATM specifications allow the switches to increase or decrease the rate of a source, all ATM switch implementations are expected to be bipolar.

## 3.6 Backward Congestion Notifications Cannot Be Used to Increase

One problem with end-to-end feedback schemes is that it may take long time for the feedback to reach the source. This is particularly true if the flow of RM cells has not been established in both directions. In such cases, switches can optionally generate their own RM cell and send it directly back to the source. The OSU scheme research showed that indiscriminate use of BECNs can cause problems. For example, consider the case shown in Figure 1. The source is sending at 155 Mbps and sends a RM cell. The switch happens to be underloaded at that time and so lets the first RM cell (C1) go unchanged. By the time the second RM cell (C2) arrives, the switch is loaded by a factor of 2 and sends a BECN to the source to come down to 77.5 Mbps. A little later C1 returns asking the source to change to 155 Mbps. The RM cells are received out of order rendering the BECN ineffective. To ensure correct operation of the BECN option, we established a set of rules. These rules are described later in Section 4.3. The first two of the six rules described there are now part of the TM specifications.

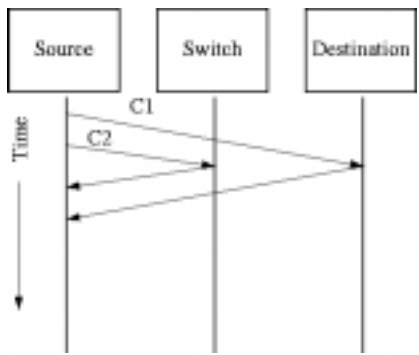

Figure 1: Space time diagram showing out-of-order feedback with BECN

# 4 Extensions of The OSU Scheme

## 4.1 Aggressive Fairness Option

In the basic OSU scheme, when a link is outside the TUB, all input rates are adjusted simply by the load level. For example, if the load is 200%, all sources will be asked to halve their rates regardless of their relative magnitude. This is because our goal is to get into the efficient operation region as soon as possible without worrying about fairness. The fairness is achieved after the link is in the TUB.

Alternatively, we could attempt to take steps towards fairness by taking into account the current rate of the source even outside the TUB. However, one has to be careful. For example, when a link is underloaded there is no point in preventing a source from increasing simply because it is using more than its fair share. We cannot be sure that underloading sources can use the extra bandwidth and if we don't give it to an overloading (over the fair share) source, the extra bandwidth may go unused.

The aggressive fairness option is based on a number of considerations. These considerations or heuristics improve fairness while improving efficiency. However, these heuristics do not guarantee convergence to fair operation. We will hence use them outside the TUB, and the TUB algorithm inside the TUB.

The considerations for increase are:

1. When a link is underloaded, all its users will be asked to increase. No one will be asked to decrease.

2. The amount of increase can be different for different sources and can depend upon their relative usage of the link.

3. The maximum allowed adjustment factor should be less than or equal to the current load level. For example, if the current load level is 50%, no source can be allowed to increase by more than a factor of 2 (which is equivalent to a load adjustment factor of 0.5).

4. The load adjustment factor should be a continuous function of the input rate. Any discontinuities will cause undesirable oscillations and impulses. For example, suppose there is a discontinuity in the curve when the input rate is 50 Mbps. Sources transmitting 50-$\delta$ Mbps (for a small $\delta$) will get very different feedback than those transmitting at 50+$\delta$ Mbps.

5. The load adjustment factor should be a monotonically non-decreasing function of the input rate. Again, this prevents undesirable oscillations. For example, suppose the function is not monotonic but has a peak at 50 Mbps. The sources transmitting at 50+$\delta$ Mbps will be asked to increase more than those at 50 Mbps.

The corresponding considerations for overload are similar to the above.

As noted, these heuristics do not guarantee convergence to fairness. To guarantee fairness in the TUB, we violate all of these heuristics except monotonicity.

A sample pair of increase and decrease functions that satisfy the above criteria are shown in Figure 2. The load adjustment factor is shown as a function of the input rate. To explain this graph, let us first consider the increase function shown in Figure 2(a). If current load level is $z$, and the fair share is $s$, all sources with input rates below $zs$ are asked to increase by $z$. Those between $zs$ and $z$ are asked to increase by an amount between z and 1.

Figure 2(b) shows the corresponding decrease function to be used when the load level $z$ is greater than 1. The underloading sources (input rate $x <$ fair share) are not decreased. Those between $s$ and $zs$ are decreased by a linearly increasing factor between 1 and $z$. Those with rates between $zs$ and $c$ are decreased by the load level $z$. Those above $c$ are decreased even more. Notice that when the load level $z$ is 1, that is, the system is operating exactly at capacity, both the increase and decrease functions are identical (a horizontal line at load reduction factor of 1). This is important and ensures that the load adjustment factor is a continuous function of $z$. In designing the above function we used linear functions. However, this is not necessary. Any

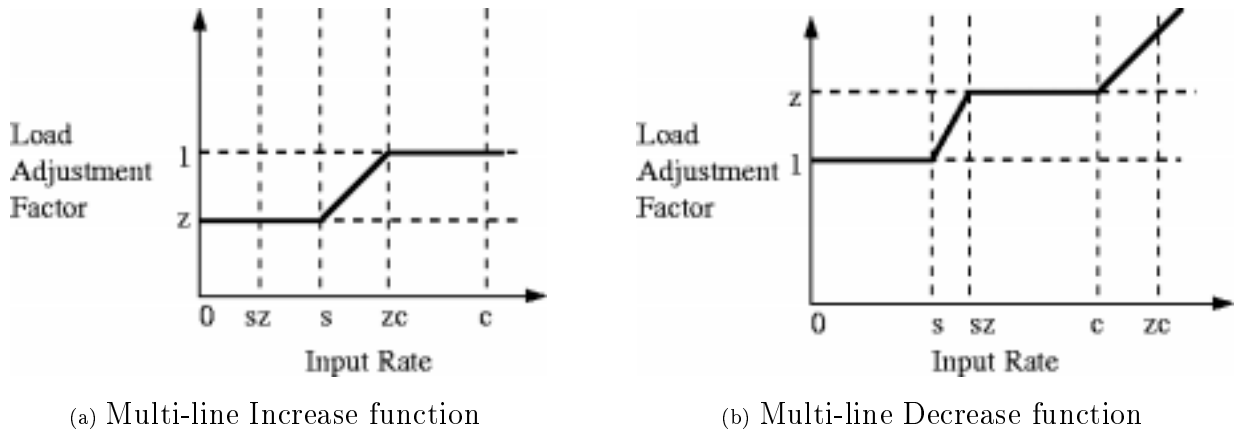

<sub>(a)</sub> Multi-line Increase function  <sub>(b)</sub> Multi-line Decrease function

Figure 2: Multi-line Increase and Decrease Functions

increasing function in place of sloping linear segments can be used. The linear functions are easy to compute and provide the continuity property that we seek.

The detailed pseudo code of aggressive fairness option is given in appendix B.

### 4.2 Precise Fair Share Computation Option

Given the actual rates of all active sources, we can exactly calculate the fair share using the MIT algorithm [1] (MIT scheme uses desired rates). Thus, instead of using only the number of active VCs, we could use the OCRs of various sources to compute the fair share. This option yields a performance much better than that possible with MIT scheme because of the following features that are absent in the MIT scheme:

1. Provide a bipolar feedback. The switches can increase as well decrease the rate in the RM cell. This avoids the extra round trip required for increase in the MIT scheme.

2. Measure the offered average cell rate at the source and use it also to compute the fair share. Using measured value is better than using desired rates.

The detailed pseudo code of precise fair share computation is given in appendix B.

### 4.3 BECN Option

For long-delay paths, backward explicit congestion notifications (BECNs) may help reduce the feedback delay. Experiments with BECNs showed that, BECNs may cause problems unless handled carefully. In particular, we established the following rules for correct operation of the BECN option with OSU scheme:

1. The BECN should be sent only when a switch is overloaded **AND** the switch wants to decrease the rate below that obtained using the LAF field of the RM cell. There is no need to send BECN if the switch is underloaded.

2. The RM cell contains a bit called "BECN bit." This bit is initialized to zero at the source and is set by the congested switch in the BECN cell. The cells that complete the entire path before returning to the source are called forward explicit congestion notification (FECN) cells. They have the bit cleared.

3. All RM cells complete a round-trip. The switch which wants to send a BECN waits until it receives an RM cell, makes two copies of it and sends one copy in the forward direction. The other, called the "BECN cell," is sent back to the source.

4. The RM cell contains a timestamp field which is initialized by the source to the time when the RM cell was generated. The timestamp is ignored everywhere except at the source.

5. The source remembers the timestamp of the last BECN or FECN cell that it has acted upon in a variable called "Time already acted (Taa)." If the timestamp in an returned RM (BECN or FECN) cell is <u>less</u> than Taa, the cell is ignored. This rule helps avoid out-of-order RM cells.

6. If the timestamp of an RM cell received at the source is equal to or greater than Taa, the variable New TCR is computed as in section 2.1. In addition, if the BECN bit is set, we ignore the feedback if it directs a rate increase :

$$\text{IF BECN.bit AND (TCR < New TCR) THEN Ignore}$$

The rate increase has to wait until the corresponding FECN cell returns. BECN is therefore useful only for decrease on long feedback paths.

The ATM forum has adopted the first two of the above rules. The RM cells as specified in the ATM Forum TM specifications do not contain the timestamps and the last three rules are not relevant to them. These are specific to the OSU scheme. The detailed pseudo code of BECN option is given in appendix B.

## 5  Simulation Results

The OSU scheme has been extensively simulated and the results have been presented in ATM Forum Contributions. A complete set of simulations can be found in [9]. Here we present one sample result for the basic scheme and its extensions.

This configuration consists of four VCs and three switches as shown in Figure 3. The second link is shared by VC3 and VC4. However, because of the first link, VC3 is limited to a throughput of 1/3 the link rate. VC4 should, therefore, get 2/3 of the second link. This configuration helps verify whether the scheme allocates all unused capacity to those sources that can use it. Figure 4 shows the simulation results for this configuration. In particular, the TCR for VC3 and VC4 are shown.

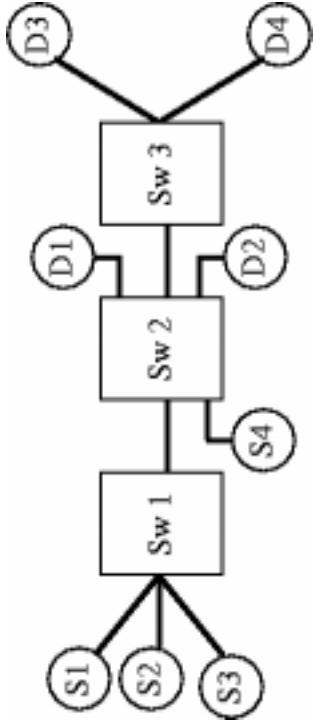

Figure 3: Network configuration with upstream bottleneck.

Notice that VC4 does get the remaining bandwidth in all cases. The top two graphs are for the LAN configuration comparing the precise fair share and the basic algorithm (indicated as single-line in the figure), for different initial rates. In each case, the rates converge to the optimum values. Observe that with precise fair share scheme, the oscillations in steady state are absent, while the single-line algorithm restricts the oscillations to the TUB. The bottom two graphs illustrate the effect of BECN for faster feedback.

## 6  Current TM Specifications vs OSU Scheme

Section 3 listed several features of the OSU scheme that have either been adopted in the standard or have been commonly implemented. In this section, we describe two features that were not adopted.

In the OSU scheme, the sources send RM cells every $T$ microseconds. This is the time-based approach. A count-based alternative is to send RM cells after every $n$ data cells. We argued that the time-based approach is more general. It provides the same feedback delay for all link speeds and source rates.

The ATM forum has adopted the count-based approach mainly because it guarantees that the overhead caused by RM cells will be a fixed percentage $(100/n)\%$ of the total load on the network.

The disadvantage with the count-based approach is that if there are many low-rate sources, it will take a long time to control them since the inter-RM cell times will be large. The time-based approach uses a fixed bandwidth per active source for RM cell overhead. For many active sources, this could be excessive.

The RM cells in the OSU scheme contain an averaging interval field. The network manager sets the averaging interval parameter for each switch. The maximum of the averaging interval along a path is returned in the RM cell. This is the interval that the source uses to send the RM cells. With the count-based approach, this field is not required.

Another major difference is the indication of rate. The OSU scheme requires sources to present both average and peak rates (along with the averaging interval) in the RM cell. The standard requires only one rate.

The OSU scheme is, therefore, incompatible with the ATM forum's current traffic management. Although, it cannot be used directly, most of its features and results can be ported to design compatible schemes. One such upgrade, called Explicit Rate Indication for Congestion Avoidance (ERICA) [11] has since been developed, which is also mentioned in the ATM Traffic Management 4.0 standards.

# 7 Limitations and Summary

This paper describes an explicit rate based congestion avoidance scheme for ATM networks. The scheme was developed as the ATM Forum traffic management specifications were being developed. While the strengths of the OSU scheme are its choice of congestion indicator, metric, and O(1) complexity, its limitations are slow convergence for complex configurations, and slight sensitivity to the averaging interval parameter. The following statements apply to the basic OSU scheme.

Our proof in appendix A is applicable to the bottleneck link (link with the highest utilization) which is shared by unconstrained sources (which can use any given allocation). It assumes that feedback is given to sources instantaneously and synchronously. In the general case, where these assumptions do not hold, the system may take longer to converge to the fair and efficient operating point. If the perturbations to the system (due to VBR, asynchronous feedback, multiple bottlenecks, or rapid changes in source load pattern) are of a time scale smaller than this convergence time, the system may be unstable. This statement is true for the convergence of *any* switch algorithm.

Further, since the scheme is measurement-based, it is slightly sensitive to the averaging interval in the switch. For example, if the number of sources is underestimated, the scheme will attempt to converge to a higher fairshare value and keep moving in and out of the TUB. Note that even then, the bottleneck is maintained at a high utilization level and the excess capacity is used to drain out queues. The number of sources is never overestimated; hence our scheme always achieves efficiency. The second quantity measured in the averaging interval is the current load level, $z$. If the system is actually overloaded, then the overload is measured correctly in $z$. However, if the system is underloaded, the averaging interval may not be long enough to exactly measure the underload. In such a case, $z$ may be underestimated, and the system may initially move to an overload region before converging.

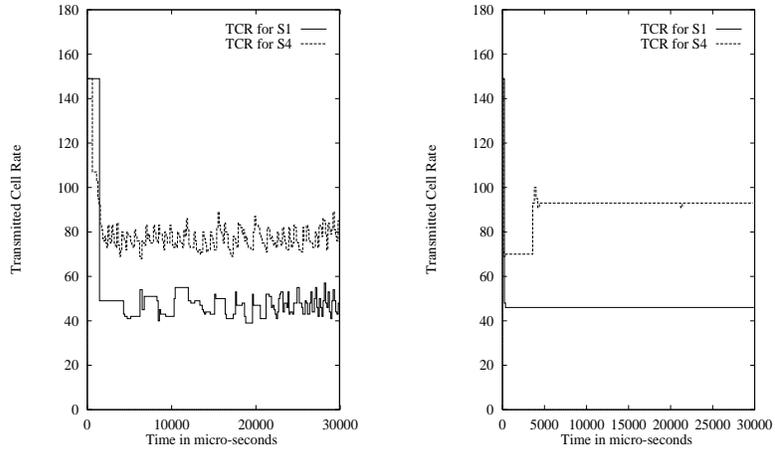
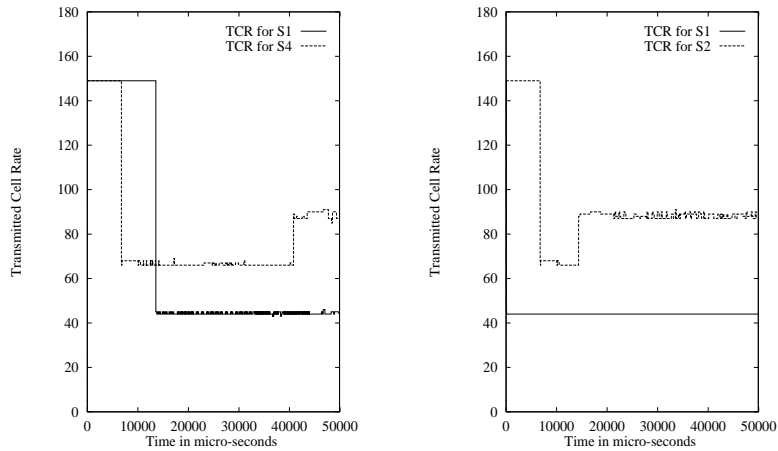

Figure 4: Simulation results for the Upstream Bottleneck configuration

Although the scheme itself is no longer strictly compatible with the specifications, many of the results obtained during this research have affected the direction of the specifications. Many features of the scheme are now being commonly used in many switch implementations. A patent on the inventions of this scheme is also pending [10].

Three different options that further improve the performance over the basic scheme are also described. These allow the fairness to be achieved quickly, oscillations to be minimized, and feedback delay to be reduced.

As stated in the previous section, we have developed a new ATM standards compatible algorithm called ERICA. ERICA and its extensions use a new set of algorithms. These algorithms achieve fast convergence and robustness for complex workloads, where input load and capacity may fluctuate arbitrarily. This will be the subject of our future publications.

## Acknowledgements

We are grateful to Rohit Goyal and Sonia Fahmy who read and helped revise the paper.

---

[3] Throughout this section, AF-TM refers to ATM Forum Traffic Management sub-working group contributions.

[4] All our papers and ATM Forum contributions are available through http://www.cis.ohio-state.edu/~jain/

## A  Proof: Fairness Algorithm Improves Fairness

In this appendix we analytically prove two claims about the simple fairness (TUB) algorithm:

**C1.** Once inside TUB, the fairness algorithm keeps the link in TUB.

**C2.** With the fairness algorithm, the link converges towards fair operation.

Our proof methodology is similar to that used in Chiu and Jain (1989)[3], where it was proven that multiplicative decrease and additive increase are necessary and sufficient for achieving efficiency and fairness for the DECbit scheme.

Consider two sources sharing a link of <u>unit</u> bandwidth. Let

| | | |
|---|---|---|
| $x$ | = | Input rate of source 1 |
| $y$ | = | input rate of source 2 |
| $z$ | = | Load level of the link $= x + y$ |
| $U$ | = | Target utilization |
| $\Delta$ | = | Half-width of the target utilization band |
| $s$ | = | Fair share rate $= U/2$ |

When $x + y = U$, the link is operating efficiently. This is shown graphically by the straight line marked "Efficiency line" in Figure 5(a). When $x = y$, the resource allocation is fair. This represents the straight line marked "Fairness line" in the figure. The ideal goal of the load adjustment algorithm is to bring the resource allocations from any point in the two dimensional space to the point marked "Goal" at the intersection of the efficiency and fairness line.

When the network is operating in a region close to the efficiency line, we consider the network to be operating efficiently. This region is bounded by the lines corresponding to $x + y = U(1 - \Delta)$ and $x + y = U(1 + \Delta)$ are in Figure 5(a). The quadrangular region bounded by these two lines and the $x$ and $y$ axes is the efficient operation zone also called the target utilization band (TUB). The TUB is described by the four conditions: $x > 0$ and $y > 0$ and $U(1 + \Delta) \geq x + y \geq U(1 - \Delta)$ Observe that $x$ and $y$ are strictly greater than zero. The case of $x = 0$ or $y = 0$ reduces the number of sources to one.

Similarly, when the network is operating in a region close to the fairness line, we consider the network to be operating fairly. This region is bounded by the lines corresponding to $y = x(1 - \Delta)/(1 + \Delta)$ and $y = x(1 + \Delta)/(1 - \Delta)$. The quadrangular region bounded by these two lines in side the TUB is called the fairness region. This is shown in Figure 5(b). Mathematically, the conditions defining the fairness region are:

$$\frac{(1 + \Delta)}{(1 - \Delta)} x \geq y \geq \frac{(1 - \Delta)}{(1 + \Delta)} x \qquad (2)$$

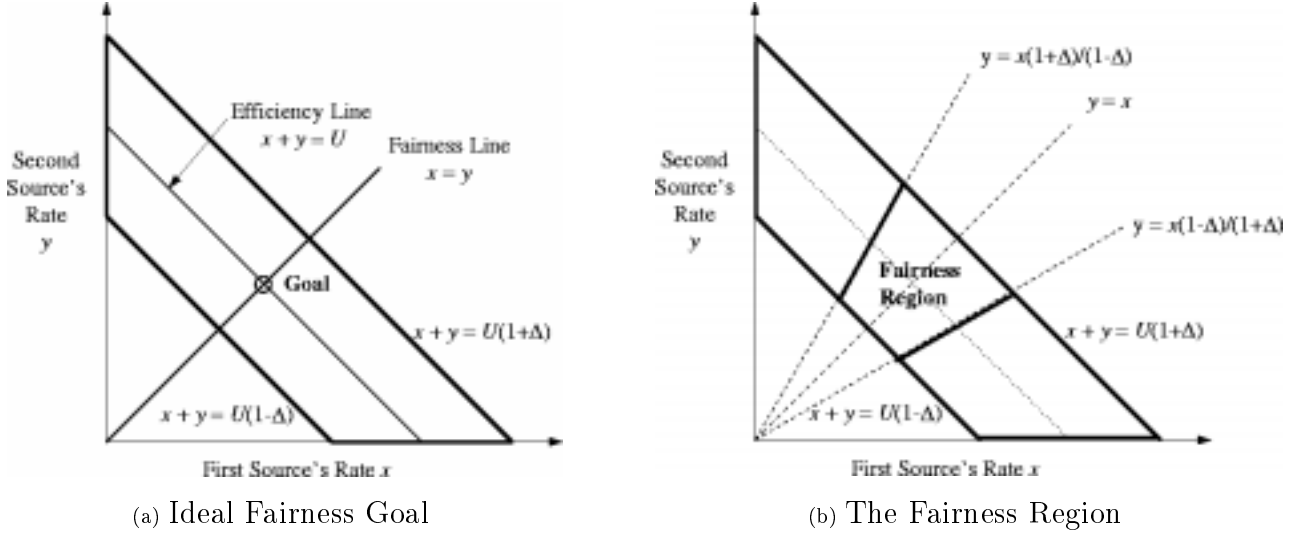

(a) Ideal Fairness Goal  (b) The Fairness Region

Figure 5: A geometric representation of efficiency and fairness for a link shared by two sources

$$U(1+\Delta) \geq x+y \geq U(1-\Delta) \qquad (3)$$

The fair share $s$ is $U/2$. Recall that the TUB algorithm sets the load adjustment factor (LAF) as follows:

IF $(x < s)$ THEN LAF = $\frac{z}{1+\Delta}$ ELSE LAF = $\frac{z}{1-\Delta}$

The rate $x$ is divided by the LAF at the source to give the new rate $x'$. In other words,
$x' = x\frac{1+\Delta}{z}$ if $x < s$ and $x\frac{1-\Delta}{z}$ otherwise.

## A.1 Proof of Claim C1

To prove claim C1, we introduce the lines $x = s$ and $y = s$ and divide the TUB into four non-overlapping regions as shown in Figure 6(a). These regions correspond to the following inequalities:

**Region 1:** $s > x > 0$ and $y \geq s$ and $U(1+\Delta) \geq x+y \geq U(1-\Delta)$

**Region 2:** $y \geq s$ and $x \geq s$ and $U(1+\Delta) \geq x+y$

**Region 3:** $s > y > 0$ and $x \geq s$ and $U(1+\Delta) \geq x+y \geq U(1-\Delta)$

**Region 4:** $y < s$ and $x < s$ and $x+y \geq U(1-\Delta)$

In general, triangular regions are described by three inequalities, quandrangular regions by four inequalities and so on.

### A.1.1 Proof for Region 1

Consider a point $(x,y)$ in the quadrangular region 1. It satisfies the conditions: $x > 0$ and $y \geq s$ and $U(1+\Delta) \geq x+y \geq U(1-\Delta)$. The link is operating at a load level $z$ given by:
$z = \frac{x+y}{U}$ or $y = Uz - x$

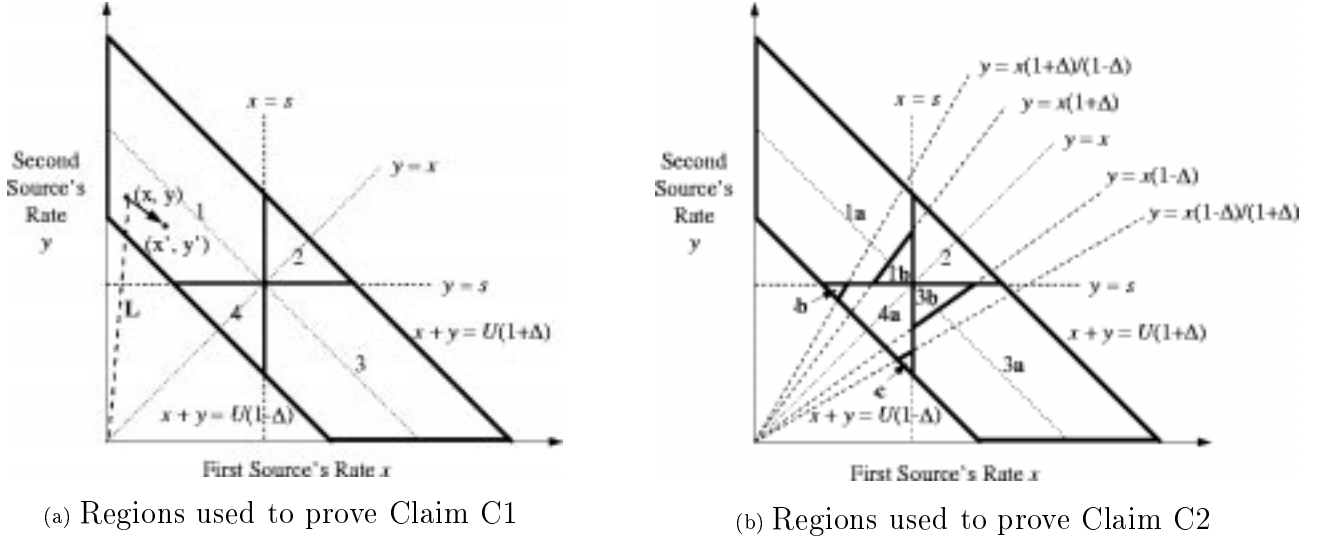

(a) Regions used to prove Claim C1

(b) Regions used to prove Claim C2

Figure 6: Subregions of the TUB used to prove Claims C1 and C2

Since $(x, y)$ is in the TUB, we have: $(1 + \Delta) \geq z \geq (1 - \Delta)$. According to the TUB algorithm, given that $x < s = U/2$ and $y \geq s = U/2$, the system will move the two sources from the point $(x, y)$ to the point $(x', y') = (\frac{x(1+\Delta)}{z}, \frac{y(1-\Delta)}{z})$.

$$x' + y' = \frac{x(1 + \Delta) + y(1 - \Delta)}{z} \tag{4}$$

$$= U(1 + \Delta) - \frac{2x\Delta}{z} \tag{5}$$

$$= U(1 - \Delta) + \frac{2\Delta}{z} y \tag{6}$$

$$\tag{7}$$

The quantity on the left hand side of the above equation is the new total load. Since the last terms of equations 5 and 6 are both positive quantities, the new total load is below $U(1 + \Delta)$ and above $U(1 - \Delta)$. In other words, the new point is in TUB. This proves that claim C1 holds for all points in region 1.

### A.1.2 Proof for Region 2

Points in the triangular region 2 satisfy the conditions: $y \geq s$, $x \geq s$, and $x + y \leq U(1 + \Delta)$

In this region, both $x$ and $y$ are greater than or equal to the fair share $s = U/2$. Therefore, the new point is given by : $(x', y') = (\frac{x(1-\Delta)}{z}, \frac{y(1-\Delta)}{z})$. Hence,

$$x' + y' = \frac{x(1 - \Delta) + y(1 - \Delta)}{z} = \frac{(x + y)(1 - \Delta)}{z} = \frac{Uz(1 - \Delta)}{z} = U(1 - \Delta)$$

This indicates that the new point is on the lower line of the TUB (which is a part of the TUB) This proves claim C1 for all points in region 2.

The proof of claim C1 for regions 3 and 4 is similar to that of regions 1 and 2, respectively.

## A.2 Proof of Claim C2

We show convergence to the fairness region (claim C2) as follows. Any point in the fairness region remains in the fairness region. Further, any point $(x, y)$ in the TUB but not in the fairness region moves towards the fairness region at every step. Consider the line L joining the point $(x, y)$ to the origin $(0, 0)$ as shown in Figure 6(a). As the angle between this line and the fairness line $(x = y)$ decreases, the operation becomes fairer. We show that in regions outside the fairness zone, the angle between the line L and the fairness line either decreases or remains the same. If the angle remains the same, the point moves to a region where the angle will decrease in the subsequent step.

We introduce four more lines to Figure 6(a). These lines correspond to $y = (1 + \Delta)x$, $y = (1 - \Delta)x$, $y = \frac{(1-\Delta)}{(1+\Delta)}x$ and $y = \frac{(1+\Delta)}{(1-\Delta)}x$. This results in the TUB being divided into eight non-overlapping regions as shown in Figure 6(b). The new regions are described by the conditions:

**Region 1a:** $s > x > 0$ and $y \geq s$ and $U(1 + \Delta) \geq x + y \geq U(1 - \Delta)$ and $y > (1 + \Delta)x$

**Region 1b:** $s > x$ and $(1 + \Delta)x \geq y \geq s$

**Region 2:** $y \geq s$ and $x \geq s$ and $U(1 + \Delta) \geq x + y$

**Region 3a:** $s > y > 0$ and $x \geq s$ and $U(1 + \Delta) \geq x + y \geq U(1 - \Delta)$ and $y < (1 - \Delta)x$

**Region 3b:** $s > y \geq (1 - \Delta)x$ and $x \geq s$

**Region 4a:** $y < s$ and $x < s$ and $x + y \geq U(1 - \Delta)$ and $y \leq \frac{(1+\Delta)}{(1-\Delta)}x$ and $y \geq \frac{(1-\Delta)}{(1+\Delta)}x$

**Region 4b:** $y < s$ and $x + y \geq U(1 - \Delta)$ and $y > \frac{(1+\Delta)}{(1-\Delta)}x$

**Region 4c:** $x < s$ and $x + y \geq U(1 - \Delta)$ and $y < \frac{(1-\Delta)}{(1+\Delta)}x$

The regions 1a and 1b are subdivisions of region 1 in Figure 6(a). Similarly, regions 3a and 3b are subdivisions of region 3, and regions 4a, 4b, and 4c are subdivisions of region 4 in Figure 6(a) respectively. Observe that regions 1b, 2, 3b and 4a are completely contained in the fairness region.

### A.2.1 Proof for Region 1a

Hexagonal region 1a is defined by the conditions: $s > x > 0$ and $y \geq s$ and $U(1 + \Delta) \geq x + y \geq U(1 - \Delta)$ and $y > (1 + \Delta)x$. The new point is given by: $(x', y') = (\frac{x(1+\Delta)}{z}, \frac{y(1-\Delta)}{z})$. Hence,

$$\frac{y'}{x'} = \frac{y}{x} \times \frac{1 - \Delta}{1 + \Delta} \tag{8}$$

Since $\Delta$ is a positive non-zero quantity, the above relation implies:

$$\frac{y'}{x'} < \frac{y}{x} \tag{9}$$

Further since $y/x$ is greater than $1 + \Delta$, equation 8 also implies:

$$\frac{y'}{x'} > (1 - \Delta) \tag{10}$$

Equation 9 says that the slope of the line joining the origin to new point $(x', y')$ is lower than that of he line joining the origin to $(x, y)$. While equation 10 says that the new point does not overshoot the fairness region. This proves Claim C2 for all points in region 1a.

### A.2.2 Proof for Region 1b

Triangular region 1b is defined by the conditions: $s > x$ and $(1 + \Delta)x \geq y \geq s$. Observe that region 1b is completely enclosed in the fairness region because it also satisfies the conditions 2 and 3 defining the fairness region.

To prove claim C2, we show that the new point given by $(x', y') = (\frac{x(1+\Delta)}{z}, \frac{y(1-\Delta)}{z})$ remains in the fairness region.

Since $(x, y)$ satisfies the conditions $1 < y/x \leq (1 + \Delta)$, we have:

$$\frac{1 - \Delta}{1 + \Delta} < \frac{y'}{x'} \leq (1 - \Delta) \tag{11}$$

Condition 11 ensures that the new point remains in the fairness region defined by conditions 2 and 3.

This proves Claim C2 for all points in region 1b.

Proof of claim C2 for region 3a and 3b is similar to that of regions 1a and 1b, respectively.

### A.2.3 Proof for Region 2

Triangular region 2 is defined by the conditions: $y \geq s$ and $x \geq s$ and $x + y \leq U(1 + \Delta)$. This region is completely enclosed in the fairness region. The new point is given by:

$$x' = \frac{x(1 - \Delta)}{z} \text{ and } y' = \frac{y(1 - \Delta)}{z}$$

Observe that:

$$\frac{y'}{x'} = \frac{y}{x} \text{ and } x' + y' = \frac{(x + y)(1 - \Delta)}{z} = U(1 - \Delta)$$

That is, the new point is at the intersection of the line joining the origin and the old point and the lower boundary of the TUB. This intersection is in the fairness region. This proves Claim C2 for all points in region 2.

### A.2.4 Proof for Region 4

Triangular region 4 is defined by the conditions: $y < s$ and $x < s$ and $x + y \geq U(1 - \Delta)$. The new point is given by:

$$x' = \frac{x(1 + \Delta)}{z} \text{ and } y' = \frac{y(1 + \Delta)}{z}$$

Observe that:

$$\frac{y'}{x'} = \frac{y}{x} \text{ and } x' + y' = \frac{(x + y)(1 + \Delta)}{z} = U(1 + \Delta)$$

That is, the new point is at the intersection of the line joining the origin and the old point and the upper boundary of the TUB.

As shown in Figure 6(b), region 4 consists of 3 parts: 4a, 4b, and 4c. All points in region 4a are inside the fairness region and remain so after the application of the TUB algorithm. All points in region 4b move to region 1a where subsequent applications of TUB algorithm will move them towards the fairness region. Similarly, all points in region 4c move to region 3a and subsequently move towards the fairness region.

This proves claim C2 for region 4.

## A.3 Proof for Asynchronous Feedback Conditions

We note that our proof has assumed the following conditions:

- Feedback is given to sources instantaneously.
- Feedback is given to sources synchronously.
- There are no input load changes (like new sources coming on) during the period of convergence
- The analysis is for the bottleneck link (link with the highest utilization).
- The link is shared by unconstrained sources (which can utilize the rate allocations).

It may be possible to relax one or more of these assumptions. However, we have not verified all possibilities. In particular, the assumption of synchronous feedback can be relaxed as shown next.

In the previous proof, we assumed that the operating point moves from $(x, y)$ to $(x', y')$. However, if only one of the sources is given feedback, the new operating point could be $(x, y')$ or $(x', y)$. This is called asynchronous feedback.

The analysis procedure is similar to the one shown in the previous sections. For example, consider region 1 of Figure 6(a). If we move from $(x, y)$ to $(x, y')$, we have:

$$y' = \frac{y(1-\Delta)}{z}$$

and

$$\begin{align}
x + y' &= \frac{xz + y(1-\Delta)}{z} \tag{12} \\
&= U(1-\Delta) + \frac{x\{z - (1-\Delta)\}}{z} \tag{13} \\
&= U(1+\Delta) - \frac{x\{(1+\Delta) - z\} + 2y\Delta}{z} \tag{14} \\
& \tag{15}
\end{align}$$

Since, the last terms of equations 13 and 14 are both positive, the new point is still in the TUB. This proves Claim C1.

Further, we have:

$$\frac{y'}{x} = \frac{y}{x}(1-\Delta)$$

Therefore,

$$\frac{y'}{x} < \frac{y}{x} \text{ and } \frac{y'}{x} \geq (1-\Delta)$$

That is, the slope of the line joining the operating point to the origin decreases but does not overshoot the fairness region.

Note that when $z = 1 - \Delta$, $y' = y$. That is, the operating point does not change. Thus, the points on the lower boundary of the TUB ( $x + y = U(1-\Delta)$ ) do not move, and hence the fairness for these points does not improve in this step. It will change only in the next step when the operating point moves from $(x, y')$ to $(x', y')$.

The proof for the case $(x', y)$ is similar. This completes the proof of C1 and C2 for region 1. The proof for region 3 is similar.

# B  Detailed Pseudocode

## B.1  The Source Algorithm

There are four events that can happen at the source adapter or Network Interface Card (NIC). These events and the action to be taken on these events are described below.

1. Initialization:
   TCR ←Initial Cell Rate;
   Averaging_Interval ←Some initial value;
   IF (BECN_Option) THEN Time_Already_Acted ←0;

2. A data cell or cell burst is received from the host.
   Enqueue the cell(s) in the output queue.

3. The inter-cell transmission timer expires.
   IF Output_Queue NOT Empty THEN dequeue the first cell and transmit;
   Increment Transmitted_Cell_Count;
   Restart Inter_Cell_Transmission_Timer;

4. The averaging interval timer expires.
   Offered_Cell_Rate ←Transmitted_Cell_Count/Averaging_Interval;
   Transmitted_Cell_Count ←0;
   Create a control cell;
   OCR_In_Cell ←Offered_Cell_Rate ;
   TCR_In_Cell ←max{TCR, OCR} ;
   Load_Adjustment_Factor ←0;
   IF (BECN_Option) THEN Time_Stamp_in_Cell ←Current Time;
   Transmit the control cell;
   Restart Averaging_Interval_Timer;

5. A control cell returned from the destination is received.
   IF ((BECN_Option AND Time_Already_Acted < Time_Stamp_In_Cell) OR
       (NOT BECN_Option))
       THEN BEGIN
           New_TCR ←TCR_In_Cell/Load_Adjustment_Factor_In_Cell;
           IF Load_Adjustment_Factor_In_Cell ≥ 1
               THEN IF New_TCR < TCR
                   THEN BEGIN
                       TCR ←New_TCR ;
                       IF(BECN_Option)
                           THEN Time_Already_Acted ←Time_Stamp_In_Cell;
                   END
               ELSE IF Load_Adjustment_Factor_In_Cell < 1
                   THEN IF New_TCR > TCR THEN TCR ←New_TCR ;
           Inter_Cell_Transmission_Time ←1/TCR;
       END; (* of FECN Cell processing *)
   Averaging_Interval ←Averaging_Interval_In_Cell;

6. A BECN control cell is received from some switch.
   IF BECN_Option

```
                THEN IF Time_Already_Acted < Time_Stamp_In_Cell
                    THEN IF Load_Adjustment_Factor_In_Cell ≥ 1
                        THEN BEGIN
                            New_TCR ←TCR_In_Cell/Load_Adjustment_Factor_In_Cell;
                            IF New_TCR < TCR
                                THEN BEGIN
                                    TCR ←New_TCR;
                                    Inter_Cell_Transmission_Time ←1/TCR;
                                    Time_Already_Acted ←Time_Stamp_In_Cell;
                                END;
                    END;
```

## B.2 The Switch Algorithm

The events at the switch and the actions to be taken on these events are as follows:

1. Initialization:
   Target_Cell_Rate ←Link_Bandwidth × Target_Utilization / Cell_Size ;
   Target_Cell_Count ←Target_Cell_Rate×Averaging_Interval;
   Received_Cell_Count ←0;
   Clear VC_Seen_Bit for all VCs;
   IF (Basic_Fairness_Option OR Aggressive_Fairness_Option )
   THEN BEGIN
       Upper_Load_Bound ←1 + Half_Width_Of_TUB;
       Lower_Load_Bound ←1 - Half_Width_Of_TUB;
   END;

2. A data cell is received.
       Increment Received_Cell_Count;
       Mark VC_Seen_Bit for the VC in the Cell;

3. The averaging interval timer expires.
       Num_Active_VCs ←max{$\sum$ VC_Seen_Bit, 1};
       Fair_Share_Rate ←Target_Cell_Rate/Num_Active_VCs;
       Load_Level ←Received_Cell_Count/Target_Cell_Count;
       Reset all VC_Seen_Bits;
       Received_Cell_Count ←0;
       Restart Averaging_Interval_Timer;

4. A control cell is received.
   IF (Basic_Fairness_Option)
   THEN IF (Load_Level ≥ Lower_Load_Bound) and (Load_Level ≤ Upper_Load_Bound)
       THEN BEGIN
           IF OCR_In_CELL > Fair_Share_Rate
           THEN Load_Adjustment_Decision ←Load_Level/Lower_Load_Bound
           ELSE Load_Adjustment_Decision ←Load_Level/Upper_Load_Bound
       END (*IF *)
       ELSE Load_Adjustment_Decision ←Load_Level;

```
IF (Aggressive_Fairness_Option)
    THEN BEGIN
        Load_Adjustment_Decision ←1;
        IF (Load_Level < Lower_Load_Bound)
            THEN IF ((OCR_In_Cell < Fair_Share_Rate×Load_Level) OR
                (Num_VC_Active =1))
                THEN Load_Adjustment_Decision ←Load_Level
                ELSE IF (OCR_In_Cell < Target_Cell_Rate×Load_Level)
                    THEN Load_Adjustment_Decision ←Load_Level + (1-
                        Load_Level)×(OCR_In_Cell/(Load_level×
                            Fair_Share)-1)/(Num_VC_Active-1)
                    ELSE Load_Adjustment_Decision ←1
            ELSE IF Load_Level ≥ Upper_Load_Bound
                THEN IF (OCR_In_Cell ≤ Fair_Share_Rate AND
                    Num_Active_VCs ≠ 1)
                    THEN Load_Adjustment_Decision ←1
                    ELSE IF (OCR_In_Cell < Fair_Share_Rate×Load_Level)
                        THEN Load_Adjustment_Decision ←max{1,
                            OCR_In_Cell/Fair_Share_Rate}
                        ELSE IF (OCR_In_Cell ≤ Target_Cell_Rate)
                            THEN Load_Adjustment_Decision ←Load_Level
                            ELSE Load_Adjustment_Decision ←
                                OCR_In_Cell×Load_Level/Target_Cell_Rate;
    END (* of Aggressive Fairness Option *)

IF (Precise_Fairshare_Computation_Option)
BEGIN
    OCR_Of_VC_In_Table ←OCR_In_Cell;
    Fair_Share_Rate ←Target_Cell_Rate/Num_VC_Active;
    REPEAT
        Num_VC_Underloading ←0 ;
        Sum_OCR_Underloading ←0 ;
        FOR each VC seen in the last interval DO
        IF (OCR_In_Cell < Fair_Share_Rate)
        THEN BEGIN
            Increment Num_VC_Underloading ;
            Sum_OCR_Underloading ←Sum_OCR_Underloading + OCR_Of_VC
        END (* IF *)
        Fair_Share_Rate ←(Target_Cell_Rate - SUM_OCR_Underloading)
            /max{1, (Num_VC_Active - Num_VC_Underloading )}
    UNTIL Fair_Share_Rate does not change (* Maximum of 2 iterations *);
Load_Adjustment_Decision ←OCR_In_Cell/Fair_Share_Rate;
END; (* Precise Fairness Computation Option *)

IF (Load_Adjustment_Decision > Load_Adjustment_Factor_In_Cell)
THEN BEGIN
    Load_Adjustment_Factor_In_Cell ←Load_Adjustment_Decision;
    IF BECN_Option and Load_Adjustment_Decision > 1
      THEN SEND_A_COPY_OF_CONTROL_CELL_BACK_TO_SOURCE ;
END (* IF *)
```